\documentclass{iau_FM}
\usepackage{graphicx}

\title[Metallicity gradients with Galactic nebular probes] %% give here short title %%
{Radial metallicity gradients with \\ Galactic nebular probes}

\author[Jorge Garc\'{\i}a-Rojas]   %% give here short author list %%
{Jorge Garc\'{\i}a-Rojas$^{1,2}$}
\affiliation{$^1$Instituto de Astrof\'\i sica de Canarias, E-38200 La Laguna, Tenerife, Spain \\ email: {\tt jogarcia@iac.es} \\[\affilskip]
$^2$Dept. de Astrof\'{\i}sica, Universidad de La Laguna, E-38206, La Laguna, Tenerife, Spain}

\pubyear{2018}
\jname{Astronomy in Focus, Volume 12} 
\editors{Piero Benvenuti, ed.}
\begin{document}

\maketitle

\begin{abstract}
The study of radial metallicity gradients in the disc of the Milky Way is a powerful tool to understand the mechanisms that have been acting in the formation and evolution of the Galactic disc. In this proceeding, I will put the eye on some problems that should be carefully addressed to obtain precise determinations of the metallicity gradients.
\end{abstract}

\firstsection % if your document starts with a section,
\section{Introduction}

It is well known that the abundance gradients in the disc of our Galaxy are more difficult to determine than in external galaxies mainly owing to distance uncertainties (especially for planetary nebulae) but also owing to the lack of objects in the inner and outer parts of the Galactic disc which is heavily obscured by the presence of dust close to the Galactic plane. Owing to Fe is strongly depleted onto dust grains in the interstellar medium (ISM), other metallicity tracers should be used, mainly O, but also $\alpha$-elements (Ne, Ar, Cl, S). In the case of H~{\sc ii} regions, O is a probe to trace the present-day chemical composition of the ISM, and in PNe, O and $\alpha$-elements are an archive of abundances in the past because in principle, these elements are not synthetised in the progenitor stars of PNe.

\section{Radial metallicity gradients }

There are many open problems with the abundance gradients of the Milky Way such as its possible temporal evolution, the existence or not of a flattening of the gradient in the outer (or in the inner) disc of the Galaxy, or the applicability of O as a reliable element to trace the metallicity in PNe. In the following I will briefly discuss some of the major sources of uncertainties in the determination of abundance gradients using spectrophotometric data of H~{\sc ii} regions and PNe, that should be addressed to try to answer some of these open questions.

When trying to compute precise abundances in photoionised nebulae, and homogeneous analysis determining physical conditions and chemical abundances from the same set of spectra is mandatory (\cite[Perinotto \& Morbidelli 2006]{perinottomorbidelli06}). Additionally, the use of appropriate lines to compute abundances is very important; as an example, computing O$^{+}$/H$^+$ ratios from the trans-auroral [O~{\sc ii}] 7320+30 lines could introduce undesired uncertainties owing to these lines could be strongly affected by telluric emission. Using physical conditions from radio recombination lines and optical or infrared lines to compute abundances can introduce systematic uncertainties owing to the different areas of the nebula covered in the different wavelength ranges. The use of appropriate atomic data is also very important; \cite[Juan de Dios \& Rodr\'{\i}guez (2017)]{juandediosrodriguez17} showed that atomic data variations could introduce differences of  0.1-0.2 dex in the derived abundances for low-density objects, but can reach or even surpass 0.6-0.8 dex at densities above 10$^4$ cm$^{-3}$, which could be very important for young and compact PNe. Finally, in spite of the big efforts made by several groups in the last years (see \cite[Stanghellini \& Haywood 2010]{stanghellinihaywood10}; \cite[Frew et al. 2016]{frewetal16}), distance determinations uncertainties are still one of the most important sources of uncertainty on the determination of the gradient, particularly for PNe. Unfortunately, precise parallaxes for PNe by the recent Gaia DR2 have only been provided for relatively close objects and, therefore do not significantly improve the scenario (\cite[Kimeswenger \& Barr\'{\i}a 2018]{kimeswengerbarria18}).     
 
{\underline{\it Planetary nebulae}}. PNe are useful tools to constrain the chemical evolution of the Milky Way as they can probe the O abundance of the ISM over a range of epochs. There are several detailed studies in the literature that have found evidences of an evolution of the gradient with time (see e.~g. \cite[Stanghellini \& Haywood 2018]{stanghellinigaywood18}, who find evidence of a steepening of the gradient with time). However, other studies (\cite[Henry et al. 2010]{henryetal10}, \cite[Maciel \& Costa 2013]{macielcosta13}) claim that owing to the observed scatter, and the uncertainties introduced by distance, age of the progenitor stars and other effects such as radial migration (\cite[Magrini et al. 2016]{magrinietal16}), it is not clear  that there has been an evolution of the gradient with time. Finally, an important point that should be taken into account is the recent discovery of the production of O in C-rich PNe at near-solar metallicities made by \cite[Delgado-Inglada et al. (2015)]{delgadoingladaetal15}. This overproduction of O (up to 0.3 dex) makes this tracer a doubtful archive of metallicities in the past if it is not taken into account.

{\underline{\it H~{\sc ii} regions}}. The radial metallicity gradient using H~{\sc ii} regions has been widely studied (see \cite[Esteban \& Garc\'{\i}a-Rojas (2018)]{estebangarciarojas18} and references therein). However, there are several open problems, as the possible flattening of the gradient in the outskirts of the Galaxy, as it has been observed in other spiral galaxies (\cite[Bresolin et al. 2012]{bresolinetal12}). A recent study of the metallicity gradient covering a large range of Galactocentric distances and making use of very deep high-quality spectrophotometric VLT and GTC data has shown that there is not a flattening of the gradient in the Galactic anticentre (\cite[Esteban et al. 2017]{estebanetal17}). Moreover, these data have also shown that the scatter at a given Galactiocentric distance is of the order of the computed uncertainties, indicating that the ISM medium is well mixed at a given distance along the Galactioc disc. Finally, the analysis of additional data have also shown a possible flattening or drop of the gradient in the inner disc (\cite[Esteban \& Garc\'{\i}a-Rojas 2018]{estebangarciarojas18}). This behaviour has been also previously found from metallicity distributions using cepheids and red giants, and in other spiral galaxies (see \cite[Esteban \& Garc\'{\i}a-Rojas 2018]{estebangarciarojas18}, and references therein).

\end{document}